**The role of twitter in the life cycle of a scientific publication**

Emily S. Darling[1,*], David Shiffman[2], Isabelle M. Côté[1], Joshua A. Drew[3]

[1]Earth to Ocean Research Group, Department of Biological Sciences, Simon Fraser University, Burnaby, BC, Canada; esdarling@gmail.com, @emilysdarling; imcote@sfu.ca, @redlipblenny

[2] Leonard and Jayne Abess Center for Ecosystem Science and Policy, University of Miami, Coral Gables, FL, USA and RJ Dunlap Marine Conservation Program, University of Miami, Miami, FL, USA; david.shiffman@gmail.com, @WhySharksMatter

[3]Department of Ecology Evolution and Environmental Biology, Columbia University, New York NY USA; jd2977@columbia.edu, @Drew_Lab

[*]Corresponding author

**Abstract**

Twitter is a micro-blogging social media platform for short messages that can have a long-term impact on how scientists create and publish ideas. We
25  investigate the usefulness of twitter in the development and distribution of scientific knowledge. At the start of the 'life cycle' of a scientific publication, twitter provides a large virtual department of colleagues that can help to rapidly generate, share and refine new ideas. As ideas become manuscripts, twitter can be used as an informal arena for the pre-review of works in progress. Finally,
30  tweeting published findings can communicate research to a broad audience of other researchers, decision makers, journalists and the general public that can amplify the scientific and social impact of publications. However, there are limitations, largely surrounding issues of intellectual property and ownership, inclusiveness and misrepresentations of science 'sound bites'. Nevertheless, we
35  believe twitter is a useful social media tool that can provide a valuable contribution to scientific publishing in the 21$^{st}$ century.

## Introduction

Social media have fundamentally changed the way people communicate ideas and information. Traditional forms of media control a one-way flow of information, from newspapers, magazines, television and radio to the public. The rise of social media technology has revolutionized the interactive sharing of ideas using online communities, networks and crowdsourcing (Thaler et al. 2012). However, the information that is transferred through social media is not limited to your Friday night plans. Today, social media go beyond personal connections to permeate professional interactions, including scientific ones.

Scientists have been harnessing the power of social media to fundamentally speed up the pace at which they are developing and sharing knowledge, both within scientific communities and with the general public (Ogden 2013). There is a growing diversity of "social ecosystems" that support the scientific and scholarly use of social media (Bar-Ilan et al. 2012). For example, scientists are using collaborative project spaces (Wikipedia, Google Docs, figshare, GitHub), blogs and microblogs (Research Blogging, twitter), online content communities (YouTube, Mendeley, CiteULike, Zotero), and professional networking sites (Facebook, Academia.edu, LinkedIn, ResearchGate) to develop new ideas and collaborations that culminate in concrete scientific outputs.

One feature of social media is that communications can be short in length and short in lifespan. The archetype of a short message service is the twitter microblogging platform (www.twitter.com) where users post short messages, called 'tweets', of less than 140 characters. These tweets can be categorized,
65    shared, sent directly to other users and linked to websites or scientific papers (e.g., Shiffman 2012; Box 1). Currently there are more than 200 million active twitter users who post over 400 million tweets per day (http://blog.twitter.com/2013/03/celebrating-twitter7.html).

70    Here, we consider both the usefulness and limitations of a bite-sized information exchange through social media, with an emphasis on twitter, during the life cycle of a scientific paper. While twitter is but one example of a microblogging tool that might one day become outdated in the ongoing evolution of social media services, we believe that the use of short messages, and the multi-directional
75    exchange of information, will continue to have a long-term impact on the development and communication of scientific knowledge. Specifically, we target our ideas to scientists, both academically young and older, who are undecided on the value and usefulness of social media.  We draw on our own experiences with tweeting to show how social media has influenced our scientific workflow in
80    marine ecology and conservation. While we chose this field because it is our own, the examples we use are broadly applicable to ecology and evolution, and to the scientific community as a whole.

85

**Box 1. Entering the twittersphere: twitter 101**

Twitter is a microblogging and social media platform that allows users to send short messages of up to 140 characters (including spaces). The first step to engaging twitter is signing up for a free account (http://www.twitter.com). This enables you to 'follow' other twitter users, which means that you subscribe to their updates and can see their messages or 'tweets' in your feed. The best way to get started is to follow someone you find interesting (N. Baron, pers. comm.). Other twitter users can also follow your messages, which means that you now have 'followers' and that your tweets are transmitted instantly to them (Fig. 1). You can tweet your own ideas and include links to websites, categorize your ideas with hashtags (#) or directly mention other people in your tweets (@). You can also 'retweet' someone else's tweet, which means that the original tweet from another user appears in your twitter stream and is broadcasted to your followers. Tweets and retweets are the core of the twitter platform that allows for the large-scale and rapid communication of ideas in a social network.

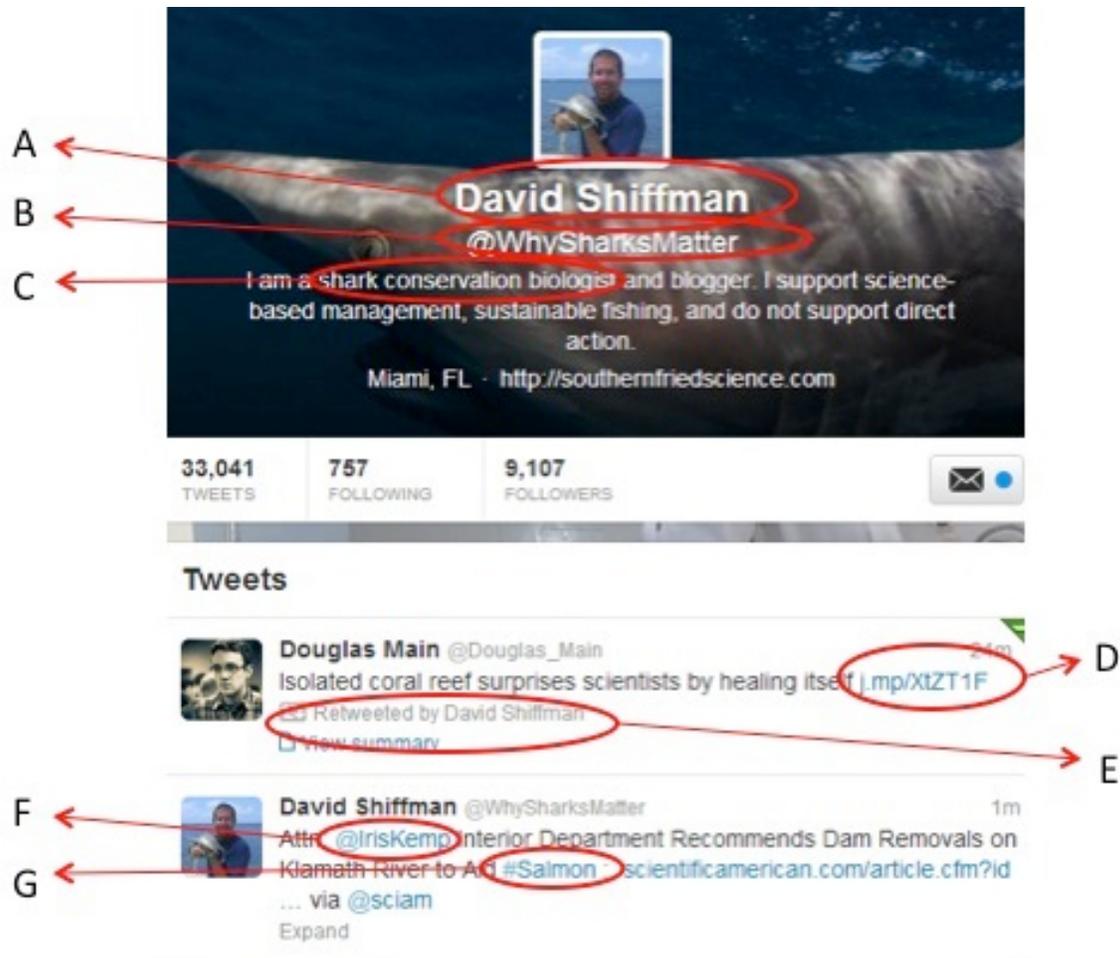

Figure 1. Examples of a twitter feed from recent tweets by the author DS who tweets as the user @WhySharksMatter. The twitter profile provides information about: A) the twitter user's real name, the name of an institution or group, or a pseudonym, B) the

105 twitter user's 'handle' or username, which is used in twitter communication with the '@' sign, C) the twitter user's description, in this case indicating that DS is a professional scientist. The twitter feed compiles a user's tweets in reverse chronological order: D) a link to a news article written about a recent science publication; E) a retweet. This tweet was originally shared by Douglas Main (@Douglas_Main, a journalist) and 're-tweeted'

110 by DS, which allows all 9,107 followers of @WhySharksMatter to see the tweet even if they do not follow Douglas Main; F) a direct mention. This is a direct mention of the user '@IrisKemp', which ensures that Iris, a graduate student studying salmon, will see the tweet regardless of whether she follows @WhySharksMatter; G) a hashtag. This highlights a hashtag, '#', which is a system of categorization within twitter. Clicking on '#Salmon' will show all other tweets that have also used this hashtag, regardless of whether you follow the user. Hashtags are a way to find and search for content that you are interested in knowing more about.

**Social media and the life cycle of a paper**

Many scientists are making the move towards social media in order to accelerate and amplify their scientific impact (Fausto et al. 2012; Fox 2012; Piwowar 2013). One in 40 scientists is active on twitter (Priem et al. 2012a), 25,000 blog entries have been indexed on the Research Blogging platform, and 2 million scientists are using Mendeley, a reference sharing tool (Piwowar 2013). Here, we consider how social media, and twitter in particular, can influence the life cycle of scientific publication, from inception and collaboration on a spark of an idea to the communication of a finished product. Specifically, we evaluate and discuss the benefits of twitter for (1) increasing scholarly connections and networks, (2) quickly developing ideas through novel collaborations and pre-review, and (3) amplifying the dissemination and discussion of scientific knowledge both within and beyond the ivory tower of academia.

*Making connections: More, faster, and interactive*

Perhaps the most obvious and important contribution of social media to scientific output is speeding up connections between scientists. Scientists have traditionally developed connections with other scientists through one-on-one interactions within their department and other local universities, and by attending professional conferences and meetings. Today, informal scholarly conversations are moving out of the 'faculty lounge' to online social media platforms, such as

140 twitter (Priem et al. 2012b, Priem 2013). The benefits of moving scholarly conversations online is that social media can provide you with a much larger "virtual department" of professional connections beyond your institution, as well as access to researchers outside of your discipline to accelerate interdisciplinary research (Bik, this issue).

145

To investigate how twitter can be used to expand a scientific network, we conducted a content analysis of twitter profiles of marine scientists who actively tweet about ocean science and conservation. We identified marine scientists based on information listed in twitter profiles and also searched publicly available

150 twitter lists of marine ecologists and evolutionary biologists (e.g., https://twitter.com/jebyrnes/eemb). We supplemented these results with marine scientists we knew to be on twitter but who were not included in these lists. Users who had pseudonymous accounts, did not identify their employer or could not be identified as professional marine ecologists were excluded. This search yielded a

155 list of 116 scientists who actively tweet about marine science and conservation, which is an extremely conservative estimate of the number of marine scientists on twitter. The majority of the 116 scientists were affiliated with universities (97, or 84%), but also included scientists from non-governmental organizations (8, 7%), government agencies (5, 4%), marine field stations (3, 2.5%) and museums or aquaria (3, 2.5%).

We compared the relative sizes of each user's traditional academic departments with the size of their virtual department of followers on twitter. We found that virtual departments on twitter were substantially larger than the average academic department (Fig. 2a). The median number of Twitter followers (241, mean ± sd: 669 ± 1600 followers, n = 116) was approximately 730 times larger than the average academic department (median: 33, mean ± sd: 37 ± 24.5 faculty; paired *t*-test, *t* = -4.08, n = 116, *p* < 0.0001, Fig. 2b).

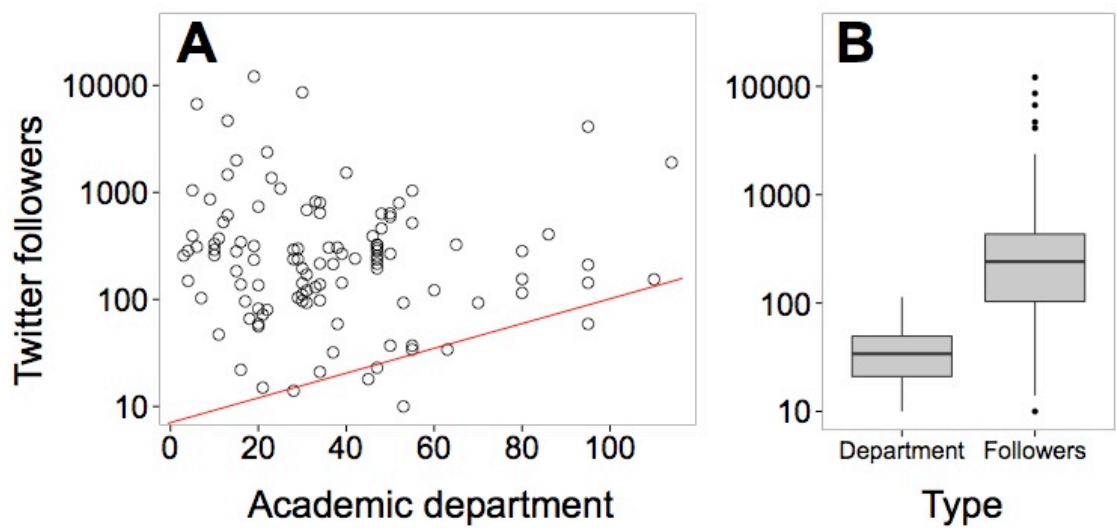

Figure 2. Twitter provides access to a 'virtual' department of followers that almost always exceeds the size of traditional academic departments. (A) The number of full-time faculty members in each scientist's institution vs the number of twitter followers for 116 marine scientists; note that the number of twitter followers is on a log scale. The red line shows the one-to-one slope. (B) The average number of twitter followers was 730 times larger

than the average number of full-time faculty members in each scientist's department. Boxplots indicate medians (thick horizontal lines), first and third quartiles (boxes), 95% confidence intervals (whiskers) and outliers (points).

180 However, not all people that follow scientists on twitter are scientists, or even scientists with whom you share common interests or wish to collaborate with. What types of people are included in the 'virtual departments' that follow scientists on twitter? We categorized the followers of each of the four authors of this paper into major groups. For ESD (n = 265 followers) and IMC (n = 285), we
185 categorized every follower, while for DS (n = 9107) and JAD (n = 1552), we categorized only the most recent 375 followers for each to make the sample sizes similar across all four authors. Followers included all types of scientists and scholarly organizations, as well as non-scientists, non-governmental organizations and media representatives (Fig. 3). The majority of our followers
190 (~55%) comprised science students, scientists or scientific organizations that could be potential collaborators for most scientists. The remaining 45% comprised non-scientists, media and the general public who may be more likely to be engaged in the dissemination of published scientific findings (see *Communicating and discussing published ideas* below).

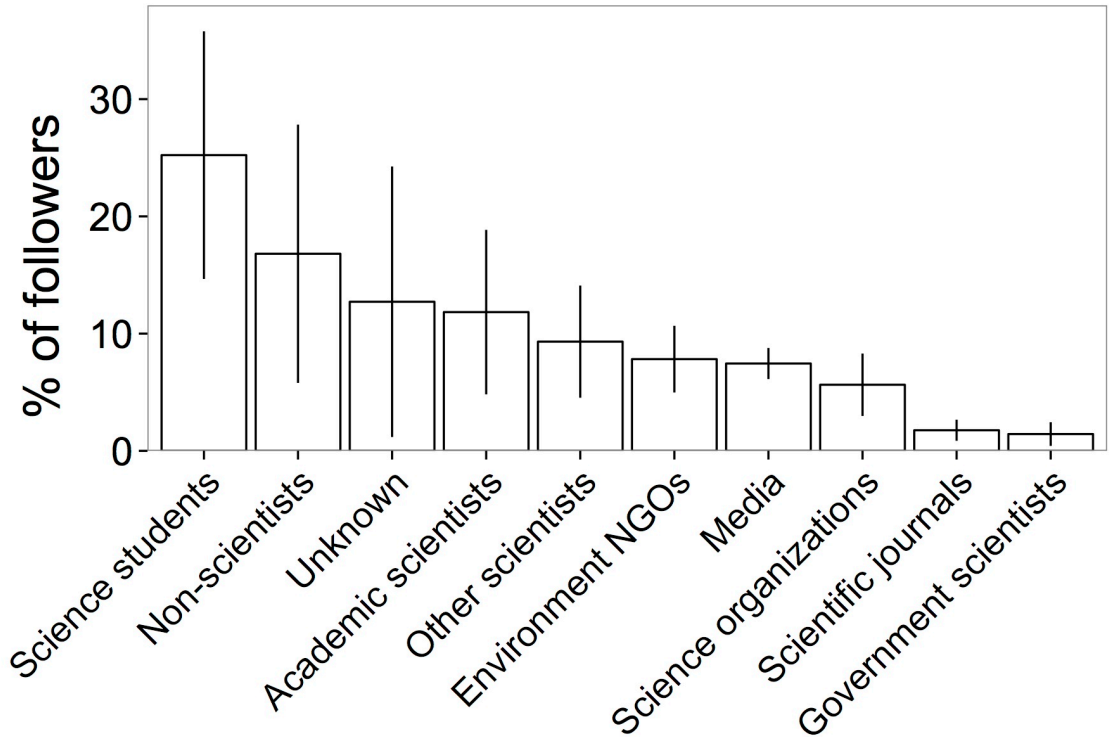

Figure 3. What types of followers do scientists have on twitter? We classified twitter followers of each of the four authors of this paper into major scholarly and non-scholarly categories. 'Science students' includes undergraduates, graduate students and postdocs. 'Other scientists' are science professionals who could not be classified into a more specific category. 'Science organizations' comprise universities, conferences, professional organizations and online science associations. The 'unknown' category is for users who did not provide information on their profile that allowed them to be classified into another group. Bars indicate mean values with standard deviation.

Finally, twitter can also be used to build and engage networks at scientific meetings and conferences. Following 'live tweets' from conferences is another strategy for building your own virtual department of colleagues on social media. Conference live-tweeting is simply sharing the information presented at a scientific conference in real-time via twitter (Shiffman 2012). This allows scientists who are not attending a meeting (or who are attending another session at the meeting) to be able to follow and, to some extent, participate in the discussion surrounding a presentation. For example, 1731 tweets were sent by 176 delegates at the 2011 International Congress for Conservation Biology. Given the cumulative number of people following these delegates, at least 110,000 twitter users could have seen tweets from this conference, an audience far larger than the one in attendance. Most ecology and evolution conferences now have hashtags (see Figure 1) that allow you to follow content, connect with new colleagues and stay on the front lines of your field.

In today's age of big data and even bigger global-scale issues, many ecologists and evolutionary biologists are conducting research that requires international and interdisciplinary collaborations (Hampton et al. 2013). Social media are a tool in your arsenal that can allow for fast and short communications that can increase and accelerate your scholarly interactions. From our own experiences, we have found that twitter enormously speeds up new connections with other scientists, which can lead to the development of new collaborations and scientific outputs, such as the commentary you are reading now.

*Moving ideas forward: open science in real time*

Social media and twitter can rapidly increase your connections to like-minded researchers, both within your field and outside it (Bik, this issue). This can lead to the next step of the scientific life cycle: turning your ideas into a scientific output. Social media won't help you write a manuscript (in fact, turning off social media might be your best bet at that point!). Nevertheless, rapid communications using social media can provide a novel arena to quickly develop and pre-review scientific ideas before submitting the final product to a peer-reviewed journal.

A recent exchange on twitter illustrates our point (Fig. 4). On 10 February 2013, Dr John Bruno (@JohnFBruno) tweeted a figure showing that the composition of coral reef fish in Belize had changed after the invasion by predatory Indo-Pacific lionfish (Fig. 4A). Within three days, there were 20 responses to his tweet; the full correspondence can be viewed here: https://twitter.com/JohnFBruno/status/300592645967859713. One reply was from John Sexton (@diverdutch), a recreational SCUBA diver who wanted to know more about the research finding in less jargon (Fig. 4B). Another reply was from Grantly Galland (@GrantlyG), a PhD student at the Scripps Institute of Oceanography, who had previously conducted fish surveys in the same area (Fig. 4C). A series of tweets between Dr Bruno, his student Abel Valdivia, and Mr Galland opened the door for a future collaboration.

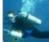

Figure 4. An example of how tweeting can move ideas forward and contribute to scientific outputs.  A) Initial tweet by Dr John Bruno, offering a new, unpublished finding for comment on twitter; B) a reply from a non-scientist, asking for non-technical explanation; C) a reply from scientist Grantly Galland, who has relevant data to contribute to the project.  The scientists strike a new collaboration.

Other scientific disciplines commonly use online resources for the pre-review of ideas and draft manuscripts. For example, economists commonly use "working papers" (i.e., draft manuscripts) and blogs to develop and share ideas before they are subject to peer-review (Fox 2012). In mathematics and physics, draft manuscripts (for example this one) are routinely submitted to the online preprint server arXiv (www.arXiv.org), which allows other scientists to view and keep up to date with the most recent research (Schriger et al. 2011, Shuai et al. 2012). However, researchers in ecology and evolution have been slower to adopt these practices. Why? Fox (2012) suggests that there is not a long-standing culture of sharing unpublished work and that there is a general lack of incentives to use of social media. For example, tweeting and blogging are rarely valued by hiring, tenure and promotion committees or granting agencies (Mandavelli 2011, Fox 2012), although this is starting to change (Piwowar 2013, Priem 2013). Ecologists and evolutionary biologists might also lack dedicated online platforms to discuss work-in-progress (although arXiv is open to these fields). We suggest that social media can contribute to this process as an open access and free service to crowdsource expert and non-expert opinions on work in progress (Fausto et al. 2012). We also envision that many different social media services could be used in this workflow, such as twitter for quickly broadcasting information and links to content, figshare (www.figshare.org) for sharing figures or presentations, and blogs for more substantive discussions (Mandavelli 2011, Fausto et al. 2012, Fox 2012).

One concern about sharing ideas on social media is that they might be 'scooped' before they are published in peer-reviewed journals. This is a possibility. However, tweeting or blogging ideas and new analyses can provide a 'time stamp' for ideas that are yours (Ogden 2013, see Fig. 5). Ideas shared as tweets are dated and searchable in the twitter archive while images or presentations uploaded onto figshare are given a citeable digital object identifier (DOI) number. These social media 'time stamps' are a way to mark and share your work without the often prolonged wait times of the traditional peer-review process. Social media are at the frontier of sharing new ideas and there will undoubtedly be different opinions among users about how these tools should be used. Nevertheless, we believe these tools have a valuable role to play in turning ideas into publications during the scientific workflow.

***Communicating and discussing published ideas***

The final step in the scientific life cycle is communicating the findings of your scientific publication to those who need to see it. Twitter can help you do this.

*Increasing reach*

Passive dissemination, which occurs when authors simply hope that their work will be discovered through tables of content in journals or serendipitous browsing, is a poor strategy. Active dissemination obviously requires more time, effort and connections, but social media (and twitter in particular) can greatly facilitate this task. The two main advantages of tweeting in terms of dissemination are that tweets allow you to convey the most interesting discoveries or conclusions of a new paper both more informally and more informatively than a paper's title can. These advantages are even being taken into consideration by some journals (e.g., Methods in Ecology and Evolution, Journal of Ecology) and conferences (the 2013 International Congress for Conservation Biology), which require each 310  submission to be accompanied by a tweetable abstract. In addition, just as twitter provides a large virtual audience for the development of ideas, it also provides an echo chamber for the dissemination of published papers. Priem and Costello (2010) estimated that approximately one-third of tweets sent by academics from the sciences, social sciences and humanities contain a hyperlink to a peer-
315  reviewed resource (e.g., a pdf - either theirs or someone else's). These so-called 'citation tweets' may be short-lived and reach only the user's followers, but the size of the audience may increase exponentially if they are retweeted. Priem and Costello (2010) found that 19% of links to peer-reviewed articles sent by a small sample of academics were retweets. In contrast, nearly half (47%) of tweets sent
320  by Nature Chemistry (2013) were retweeted, on average 4 times each. Admittedly, only about one-third of Nature Chemistry's tweets had links (to papers, but also to blogs and other chemistry-related sites), and it is not clear whether tweets with links are more or less likely to be retweeted than those without. Nevertheless, a citation tweet that is subsequently retweeted can reach
325  an immensely wide audience, with relatively little effort on the part of the initial author. Sharing published work can also restart the scientific life cycle if another researcher follows up on an idea or forms a new collaboration based on a citation tweet.

330  Because the followers of scientists comprise not just academics, their citation tweets (and retweets) will reach eyes much beyond the ivory tower, into non-governmental organizations, private industry, government agencies and non-scientists (see Fig.3). This means that science with applied or policy implications can reach people in decision-making positions. Tweeting directly to a decision maker (e.g., by sending a tweet to @barackobama or @pmharper; see Fig. 1) also makes it possible to reach such people, even if they are not your followers. In the United States, all of the members of the House of Representatives (https://twitter.com/cspan/u-s-representatives/members) and the Senate (https://twitter.com/gov/us-senate/members) have a twitter account (updated from Lassen & Brown 2011) as do more than three-quarters of members of the federal parliament in Canada (http://poliTwitter.ca/page/canadian-politicians-on-Twitter; accessed 22 March 2013). Of course, there is no guarantee that a politician will click on a link sent by a scientist.

However, journalists might pick up on tweets about new science. For example, a paper about the natural history of the cookiecutter shark (*Isistius brasiliensis*; Hoyos-Padilla et al. 2013) had been published online for three months before it was discovered by one of us (DS). After DS tweeted the key points and a link to the paper, he was contacted by several science journalists who follow him on twitter. One of these journalists was a writer for Our Amazing Planet, Douglas Main (@DouglasMain), who wrote the first story about this scientific article (http://news.yahoo.com/cookiecutter-shark-takes-bite-great-white-142359513.html). Following this media attention, the story was picked up by National Geographic (http://phenomena.nationalgeographic.com/2013/01/23/what-bit-this-great-white-

shark-a-cookie-cutter/), and several news outlets. Conference live tweets can also expose your science to journalists. At the Ecological Society of America meetings in August 2012, one of us (JAD) tweeted about his upcoming talk on shark weapons and shifting baselines in the Pacific Ocean. Ed Yong (@edyong209), a journalist who follows JAD (@Drew_Lab), saw the tweet, attended the talk and then wrote a blog about it for Discover Magazine (http://blogs.discovermagazine.com/notrocketscience/2012/08/13/weapons-made-from-shark-teeth-are-completely-badass-and-hint-at-lost-shark-diversity/#.UV2pFaugl4r) and Nature News (http://www.nature.com/news/shark-tooth-weapons-reveal-lost-biodiversity-1.11160). Using social media to build a network of journalists can be an excellent tool to promote the popularization of scientific findings.

Finally, because of their great potential for wide dissemination, tweets might eventually come to complement, or even replace, the usual outputs of workshops and conferences, such as consensus papers, proceedings or books. One of us (IMC) recently convinced the organizers of an upcoming conference on marine conservation to propose a tweetable vision statement as the main output for the meeting. These 140 characters can then be tweeted to relevant policy-makers and politicians, and retweeted to people far beyond the reach of a conventional output.

*Increasing impact*

380 The impact of scientific papers has traditionally been measured in terms of numbers of citations (Neylon and Wu 2009). Tweeting can influence this impact metric. For example, articles published in the Journal of Medical Internet Research that were tweeted about frequently in the first three days following publication were 11 times more likely to be highly cited 17 to 29 months later than

385 less tweeted articles (Eysenbach 2011). In fact, top-cited articles could be predicted quite accurately from their early tweeting frequency (Eysenbach 2011). In a separate study of ~4600 scientific articles published in the preprint database [arXiv.org](arXiv.org), Shuai et al. (2012) found that papers with more mentions on twitter were also associated with more downloads and early citations of papers,

390 although the causality of these relationships is unclear (Shuai et al. 2012).

The importance of a scientific paper should arguably capture more than its immediate citation within the academic community, but also its use and impact beyond academia as well as the range of alternative communities reached.

395 For these reasons, it is becoming increasingly clear that number of citations is an unduly narrow way to measure scientific impact (Neylon and Wu 2009, Priem 2013). Alternative metrics (or 'altmetrics') that quantify the broader impact and reach of scientific knowledge beyond traditional journals are being actively developed (e.g., Eysenbach 2011, Priem et al. 2012b; Priem 2013). Many are

400 based on data derived from sharing and social media, including the volume of downloads for a paper or data repository (such as Dryad, figshare), the number of mentions on sites like Facebook and twitter, and bookmarks to online referencing libraries like Mendeley or CiteULike (Piwowar 2013). There are also third-party platforms, such as ImpactStory.org and Altmetric.com, that help researchers measure the online impact of their science (Piwowar 2013; Priem 2013).

The use of altmetrics is still controversial in the scientific community. Some scientists are concerned that altmetrics are largely untested indicators that can be influenced by fraud and cheating, such as automating paper downloads from multiple fake user accounts or 'robot tweeting' (Cheung 2013). However, 'gaming' the algorithms used for creating altmetrics may be more difficult than it seems and there are emerging scholarly defenses against cheats or hacks to the system (Priem 2013). Furthermore, altmetrics are not intended to replace traditional bibliometrics like number of citations – in fact, these two approaches are complementary and capture different types of impact for different audiences (Priem et al. 2012b). As tenure committees and funding agencies begin to demand science that informs policy or provides meaningful change and demonstrated outcomes, altmetrics may change the playing field of how we recognize and reward scientific outputs (Ogden 2013, Piwowar 2013, Priem 2013).

*Providing post-publication critiques*

Many journals, particularly open-access ones, have tried to promote online discussion of published results (Neylon and Wu 2009). Online commenting offers a potentially quicker and more informal way to comment on published work than traditional printed letters to the editor. However, for a variety of reasons, scientists have largely failed to engage in this type of post-publication critique. For example, less than 20% of articles in high-impact medical print journals that offered online commenting facilities received comments, and from 2005 to 2009, the proportion of journals offering this service declined (Schriger et al. 2011). For open-access journals in the PLoS (Public Library of Science) family, the proportion of papers with comments within three months of publication has hovered around 10-15% since 2009 (Priem 2011). In contrast, nearly one-third of articles in the British Medical Journal were criticized in the journal's 'rapid response' online commenting section, but only half of these criticisms received replies from the papers' authors (Gotzsche et al. 2010), which suggests that the goal of constructive interactions that might refine and advance published science is not fully met with the current journal-hosted commenting model.

Enter twitter, which allows rapid-fire, low-effort, pointed comments that focus on the most serious problems with a published paper. Blogs are more time-consuming (to write and to read) but are being used to the same effect (e.g., Fox 2012). The speed with which tweets and blogs questioned the validity of high-profile papers on longevity genes and arsenic-based life has been highlighted elsewhere (Mandavilli 2011). The result is a prompt weeding out of weak science, which admittedly should have occurred before publication (see *Moving ideas forward: open science in real time* above). However, the contrast between the apathy of scientists towards journal-hosted commenting and their enthusiasm for participating in twitter firing squads is striking. We believe that it is due in large part to cost effectiveness (i.e., the recognition gained from tweeting is generally low, but so is the effort expended), and perhaps to anonymity (i.e., some scientists tweet under aliases).

There is thus great value in 'trials by twitter' (*sensu* Mandavilli 2011), which can be made even greater if contributions are coherently presented. Social network services such as storify (www.storify.com) allow series of tweets to be organized (for example, chronologically or by themes) and linked by a narrative to create coherent stories or arguments. This format allows readers to assess more easily the weight of evidence presented in tweets. In fact, to encourage fruitful post-publication critique and interactions, scientific journals could appoint dedicated online tweet editors who can storify and post tweets related to their papers.

**Limitations of social media in the scientific workflow**

While twitter and other forms of social media have the potential to expedite and enhance scientific writing, they do come with a suite of potential pitfalls. Integrating a scientific workflow with social media can raise issues of intellectual property ownership, inclusiveness and misrepresentations of complex ideas as science 'sound bites'.

Most social media platforms have terms of service agreements (TOSs) that, in varying degrees, state that social media platforms retain the ability to rebroadcast the content of that media without the author's explicit permission.  This raises questions about distribution of ideas and ownership of intellectual property. For the vast majority of social media exchanges (i.e., pictures of cats), this may not be a problem. However, when these conversations turn to plans for grants, discussions of data or planning a paper, the openness and lack of control over who sees those discussions can be problematic (see *being scooped* above).

The transparency of scientific conversations on social media means that these conversations are available to a very wide audience, which often goes much beyond strictly scientific circles.  While in general this is a good thing, one can easily imagine how a frank discussion about data interpretation might be picked up and exploited to give the impression that the results are questionable or specious.  An extreme example of this was the "Climategate" scandal that began in 2009 when internal emails about the International Panel on Climate Change (IPCC) were leaked. Climate change deniers conflated scholastic debates over the interpretation of data with a complete fabrication of those results and used this deceptive argument to try to discredit the IPCC's findings (Leiserowitz et al. 2013). With twitter, everything is in the open and we urge scientists to be mindful in their exchanges. This is not to say that there should not be frank and honest scientific discussions, but we must acknowledge that those discussions may be hijacked by people with outside agendas. Twitter at times feels like an intimate dinner chat, but in reality it's more like having your conversation broadcast on national news.

An additional area of concern with reliance on social media platforms to carry out science is one of inclusiveness. As with all crowdsourcing, one must know who is doing the communicating (Ogden 2013). Currently only a small proportion of scientists tweet, estimated as 1 in 40 (Priem et al. 2012a). A common concern for scientists who are skeptical of social media is that only 'younger' people, such as graduate students and early career researchers, are using it. To test this assumption, we collected information on the 'academic age' (i.e., the number of years since receiving a PhD) of the tweeting marine scientists in our content analysis described above. Students in the process of obtaining their PhD, or other degree, were assigned an academic age of zero years. The majority (62%, 73 out of 116) of marine scientists using twitter had received their PhD within the last 5 years (Fig. 5). However, there were a handful of prominent scientists who had received their PhDs more than 20 years ago. Thus while most scientists who actively use twitter are indeed newer to academia, there are more experienced scientists who actively tweet. But nevertheless, marine scientists on twitter are a non-random sample of total marine scientists and skewed towards younger investigators and newer academics (see Fig. 3). This means that interacting with a twitter science community may involve conversations with individuals that have a different suite of experiences than the more traditional scientific community.

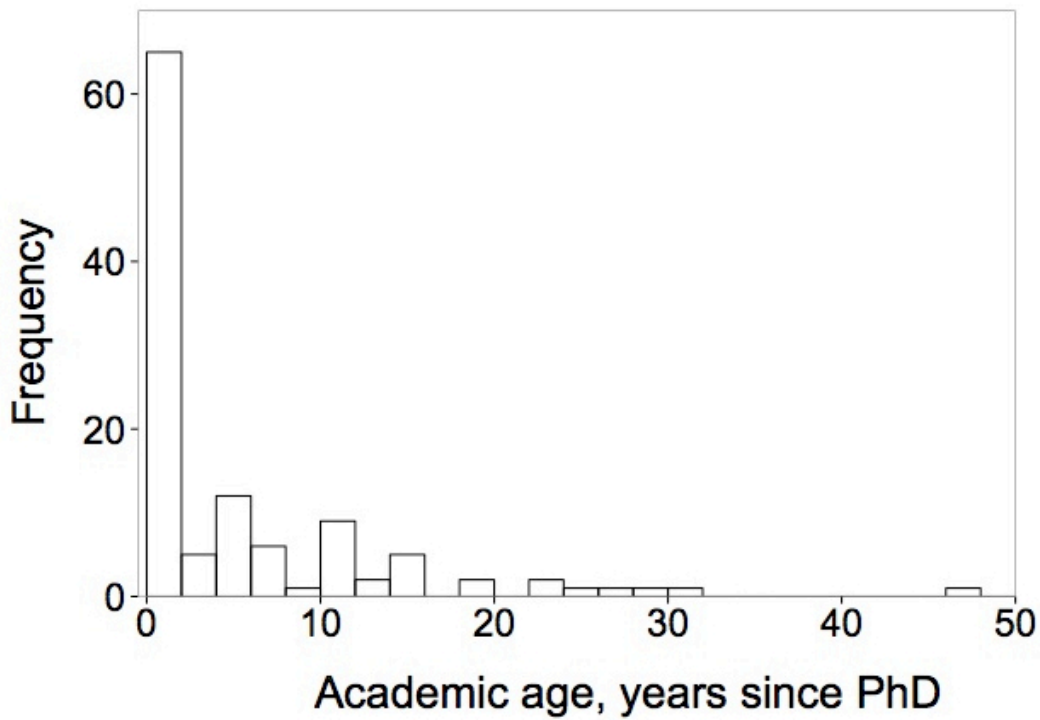

Figure 5. Age distribution (years since PhD) of 116 marine scientists who are active on twitter. The majority of scientists are academically young, having obtained their PhD degrees less than 5 years ago; however, more experienced scientists also tweet.

Lastly, the character limit imposed by twitter necessitates brevity when describing one's ideas. While we suggest that distilling ideas into tweetable lengths is an exceptionally useful skill (especially when transferred into writing page-limited grants, for example), one must use caution to not lose important nuances in a quest to fit an idea into one tweet, or 'science soundbite'. As with other written forms of communication, vocal cues are invisible in tweets and misrepresentation 530 of phrases can cause complications. Similarly there are some ideas that are too complex to shoehorn into a single tweet. In those cases we suggest either 1) breaking the idea up into successive tweets, or 2) linking to a blog or journal article where the thoughts can be explained in a more leisurely and extensive fashion.

535

**Conclusions**

Social media have changed the playing field for how scientists interact with each other and beyond the ivory tower of academia into policy and public arenas. We hope our experiences with social media, and twitter in particular, will encourage
540 hesitant scientists to give it a spin - we believe there can be great and unexpected value to including social media into the life cycle of a scientific paper. Trying new things and taking risks will be part of the future of scholarly communication and publication (Priem 2013). We're doing it and you can too – tweet us to let us know how it goes or if you have any questions along the way.

545

**Acknowledgements**

We thank Jarrett Byrnes for the invitation and encouragement to prepare this article (via twitter!). ESD and IMC were supported by the Natural Sciences and Engineering Research Council of Canada. DS is supported by the Guy Harvey
550 Ocean Foundation, the RJ Dunlap Marine Conservation Program, and the Leonard and Jayne Abess Center for Ecosystem Science and Policy. He would like to thank Dr. Andrew David Thaler (@SfriedScientist) for introducing him to twitter, and his adviser Dr. Neil Hammerschlag (@DrNeilHammer) for all his support.

555